\documentclass[aps,prl,showpacs,twocolumn]{revtex4-2}
\usepackage{graphicx}
\usepackage{subfig}
\usepackage{tabularx}
\usepackage{bm}
\usepackage{amsmath}
\usepackage{esvect}
\usepackage[capitalize]{cleveref}

\providecommand{\abs}[1]{\left\lvert#1\right\rvert}

\providecommand{\bra}[1]{\langle #1 \rvert}
\providecommand{\ket}[1]{\lvert #1 \rangle}
\providecommand{\braket}[2]{\langle #1 \rvert #2 \rangle}

\providecommand{\be}{\begin{equation}}
\providecommand{\ee}{\end{equation}}
\providecommand{\ba}{\begin{eqnarray}}
\providecommand{\ea}{\end{eqnarray}}

\usepackage{mathbbol}

\usepackage{tikz}
\usepackage{amsfonts,amssymb}
\usepackage{dsfont}
\usepackage{physics}
\usepackage{tocvsec2}

\newcommand{\beq}{\begin{equation}}
\newcommand{\eeq}{\end{equation}}

\usepackage{appendix}

\begin{document}

\title{Quantum metrology using time-frequency as quantum continuous variables: resources, sub shot-noise precision and phase space representation}

\author{Eloi Descamps$^{1,2}$}
\author{ Nicolas Fabre$^{3,4}$}
\author{ Arne Keller$^{2,5}$}
\author{Pérola Milman$^{2}$ }
\email{corresponding author: perola.milman@u-paris.fr}

\affiliation{$^{1}$Département de Physique de l’Ecole Normale Supérieure - PSL, 45 rue d’Ulm, 75230, Paris Cedex 05, France}
\affiliation{$^{2}$Université Paris Cité, CNRS, Laboratoire Matériaux et Phénomènes Quantiques, 75013 Paris, France}
\affiliation{$^{3}$ Departamento de Óptica, Facultad de Física, Universidad Complutense, 28040 Madrid, Spain}
\affiliation{$^{4}$ Telecom Paris, Institut Polytechnique de Paris, 19 Place Marguerite Perey, 91120 Palaiseau, France}
\affiliation{$^{5}$ Department de Physique, Université Paris-Saclay, 91405 Orsay Cedex, France}

\begin{abstract}

We study the role of the electromagnetic field's frequency on the precision limits of time measurements from a quantum perspective, using single photons as a paradigmatic system. We demonstrate that a quantum enhancement of precision is possible only when combining both intensity and spectral resources and, in particular, that spectral correlations enable a quadratic scaling of precision with the number of probes. We identify the general mathematical structure of non-physical states that achieve the Heisenberg limit and show how a finite spectral variance may cause a quantum-to-classical-like transition in precision scaling for pure states similar to the one observed for noisy systems. Finally, we provide a clear and consistent geometrical time-frequency phase space interpretation of our results, well identifying what should be considered as spectral classical resources.
\end{abstract}
\vskip2pc 
 
\maketitle

The wave nature of radiation make it a choice system for time precision measurements: using different interferometric techniques, precision in time is set by the inverse of the field's frequency, or of the field's bandwidth for non-monochromatic fields. Leaving apart the systematic error, classical power noise is in general a limitation, but it can be reduced to the level of the standard quantum limit (SQL) \cite{LIGO} - or shot-noise -, and scales with $1/\sqrt{\langle \hat n \rangle}$, where $\langle \hat n\rangle$ is the average photon number, or the intensity, of the field. In this case, photons behave as independent probes, which is explained by the Poissonian nature of coherent (quasi-classical) states. 

Fully exploiting  quantum resources can quadratically improve the SQL \cite{PhysRevLett.96.010401}, and in quantum optics, a sub-shot noise precision can be obtained from different field statistics corresponding to squeezed \cite{pinel_ultimate_2012, PhysRevLett.100.073601, Zhuang_2020, LIGO2}, NOON \cite{SteveGear, PRXQuantum.3.010202, PhysRevLett.99.070801} and Schr\"odinger cat-like states \cite{PhysRevA.94.022313, zurek_sub-planck_2001,LuizSub,Dalvit_2006}, for instance, as well as using non-local evolutions in multi-mode states \cite{Braun}. 

Thus, when dealing with time precision measurements, two key factors limiting the precision are usually set apart: the photon number statistics - related to the particle nature of light -, which leads to the precision scaling, and the modal properties - related to the field's wave character \cite{PhysRevX.11.031041} - which is treated as a classical ressource. However, the two aforementioned aspects of radiation are not in general independent, especially when considering intrinsically multimode non-Gaussian states. For instance in \cite{MacconeNature}, it was shown that entangled and squeezed states in frequency can lead to quantum enhanced clock synchronization and position measurement.  Nevertheless, providing a clear picture of the interplay between the two aspects of radiation in metrology and in quantum optics in general remains an open problem, in spite of its fundamental and practical importance. 

In the present Letter we address the vast problem of field and mode non-separability and its consequences on quantum metrology. By doing so, we unveil the time-frequency phase space structure behind quantum precision limits and introduce a definition of classical ressource which is common to both the modal and the particle aspects of the quantum field. The introduced geometrical picture provides a description of how scaling properties with constant resources depend on modal and particle entanglement or, in general, in the collective photonic behavior. We study in details a paradigmatic system consisting of $n$ distinguishable photons occupying each a different ancillary mode (for instance, a spatial mode) and show how quantum metrological enhancement can be obtained in such a systems. Photons in independent spatial modes are characterized by a frequency wave-function, and the frequency variable is treated quantum mechanically, since it is directly associated to each single photon's statistical properties. Notice that in this case, we consider the paraxial approximation, so that the field's transverse and longitudinal degrees of freedom factorize, as in \cite{fabre_modes_2020}. Thus, the frequency degree of freedom can be directly associated to the longitudinal mode's wave-vector.

We introduce the basic principles of quantum metrology considering the example of phase estimation. For such, we define a probe, whose evolution depends on a parameter $\theta$ to be estimated. The probe is then measured and an estimator infers the value of the parameter from the measurement results. For an unbiased estimator, the average value of $\theta$, $\langle \theta \rangle = \theta_r$, where $\theta_r$ is the {\it real} value of the parameter. The different outcomes $x$ are obtained with probability $p(x|\theta)$, and  precision is limited by the Cram\'er-Rao bound \cite{Cramer}:  $\delta \theta \geq 1/\sqrt{\nu F(\theta)}$, where $F(\theta)=\int dx \frac{1}{p(x|\theta)}\left(\frac{\partial p(x|\theta)}{\partial\theta}\right)^2 $ is the Fisher information (FI) and $\nu$ the number of independent repetitions of this procedure. In quantum metrology, the probe is a quantum state, and we can consider that it evolves by the action of a unitary operator $\hat U(\theta)$ depending on the parameter to be estimated $\theta$. The optimization of the FI over all possible measurements leads to the quantum Fisher information (QFI) $F_Q(\theta)$ \cite{QFI} that sets a more general bound for the precision of estimating the parameter $\theta$, the quantum Cram\'er-Rao (QCR) bound $\delta \theta \geq 1/\sqrt{\nu F_Q(\theta)}$.

The QFI for pure states is proportional to the overlap between the initial state and the displaced one, $F_Q(\theta) = 8(1-|\langle \psi_{\theta}|\psi_{\theta+d\theta}\rangle|)/d\theta^2 $ and $\ket{\psi_{\theta+d\theta}}= \hat U (d\theta)\ket{\psi_{\theta}}= e^{i\hat H d\theta/\hbar}\ket{\psi_{\theta}}$. Expanding the unitary operator up to second order in $d \theta$, we obtain the well known expression for the QFI, $F_Q(\theta) = 4 (\Delta \hat H)^2$ \cite{VarianceQFI}, {\it i.e.}, it is proportional to the variance of the Hamiltonian computed in the state used as a probe (the initial state). We have then an inequality that will be central to this contribution:
\beq \label{Variance}
\delta \theta \geq 1/(2\sqrt{ \nu }\Delta \hat H).
\eeq

Beating the SQL involves obtaining a scaling of the QFI better than $\langle \hat n\rangle$, which quantifies the amount of available resources (in quantum optics, the average field's intensity, in general).  Quantum mechanical scaling can be as good as the Heisenberg limit, where the QFI is proportional to $\langle \hat n ^2\rangle$ (a scaling shown to be optimal \cite{Kiok}). 

The quadrature phase space was show to provide a clear geometrical picture of \eqref{Variance}\cite{zurek_sub-planck_2001, LuizSub, PhysRevA.94.022313}, since $\hat H$ generates a phase space trajectory. The maximal precision can be seen as the minimum displacement of a Wigner function so as it becomes distinguishable from the initial one. In particular, sub-Planck structures are associated to sub-shot noise precision.

In quantum optics, phase estimation is often linked to time (or delay) estimation for single mode fields: the free evolution on different optical paths results in a phase gain proportional to the frequency of the field (which is constant for monochromatic fields).  This evolution can also be visualized as translations in the time frequency phase space (TFPS), where exotic spectral properties have also been observed but not yet associated to any quantum effect \cite{Praxmayer,WalmsleySubPlanck, fabre_parameter_2021}:  although it was shown that the right choice of modes for non-monochromatic Gaussian single-mode states is essential for optimizing precision measurements \cite{pinel_ultimate_2012, Safranek_2019}, the frequency related {\it statistical} properties of the field can be disregarded in this case and the field's spectral properties become mere quantities that do not play a role in the {\it scaling} of the QFI. Rather, they simply determine the {\it units} in which the scaling is computed (this can be seen from Eq. \eqref{OmegaNSep}, for instance). However, for many quantum states, such as intrinsically multi-mode non-Gaussian states, it may not be possible to separate the modal and the field statistics. This is the case of frequency entangled single photon states - which are the main subject of this Letter - that are used as a resource in various quantum optical protocols \cite{Qudits, PhysRevLett.103.253601,PhysRevA.82.013804, Lukens:17, PhysRevA.105.052429}, including metrological ones \cite{chen_hong-ou-mandel_2019, Jordan2, Lundeen, PhotonicQuantumMetro, lyons_attosecond-resolution_2018, MacconeNature}.  For these reasons it is crucial for quantum optical based metrological protocols to establish a consistent formalism that demonstrates how various optical resources, as modes and the field statistics, interplay and contribute to the establishment of precision limits in parameter estimation.

In order to do so, we will study free evolution as the generator of the probe state's dynamics. This will enable the definition of a common classical reference with a clear interpretation both in the quadrature phase space and in the frequency-time representation using coherent states. Then, we'll show how quantum metrological advantage can appear from frequency correlation properties and interpreted in the time-frequency phase space. Since throughout this Letter we'll mostly consider evolutions generated by Hamiltonians, we'll restrict our discussion to the variance of this operator.  

The first studied system consists of states that are separable in $n$ (orthogonal) auxiliary mode basis (as spatial modes, for instance). The free evolution Hamiltonian is given by $\hat H=\hbar \hat \Omega$, where $\hat \Omega = \sum_{i=1}^n  \alpha_i \hat \omega_i$, $\alpha_i=\pm 1$, is a collective mode operator and $i=1,...,n$ denotes the spatial modes. The operators $\hat \omega_i$ are the frequency operators (see Supplementary Material A) acting on each mode $i$. The QFI for pure states is proportional to the variance of $\hat \Omega$, which can be expressed, for mode separable states as (see Supplementary Material F):  
 \begin{equation}\label{OmegaNSep}
(\Delta \hat \Omega)^2 = \sum_{i=1}^n (\langle \hat n_i\rangle (\Delta \omega_i)^2 + (\Delta \hat n_i )^2{\bar \omega_i^2}),
\end{equation}
where $\hat n_i$ is the photon number operator in spatial mode $i$ and $\Delta \hat n_i$ its root mean square (RMS).  ${\bar \omega}_i^2 =(\int \omega |S_i(\omega)|^2 d\omega)^2$ is the average frequency squared in mode $i$. The function $S_i(\omega)$ is a complex function, or the field's spectrum in the $i$-th mode, with $\int |S_i(\omega)|^2 d\omega =1$. Thus,  $|S_i(\omega)|^2$ behaves as a classical density probability distribution. Finally, $\Delta \omega_i$ is the frequency RMS where, again, $\omega_i$ is considered as a continuous random variable with density probability distribution $|S_i(\omega)|^2$.  A first remark is that Eq. \eqref{OmegaNSep} explicits two types of contributions to the QFI: one coming from the photon number variance and another from the frequency variance. While the mechanisms by which the first one can be associated with a quantum metrological advantage have been extensively studied \cite{Science} and are related to the quadrature phase space structure, the other is often associated to a mere free classical resource, since it depends only linearly on the average photon number. 

In order to gain insight we analyze Eq.\eqref{OmegaNSep} for a coherent state of amplitude $\beta$ and spectrum $S(\omega)$ in a single spatial mode \cite{combescure_coherent_2012, PhysRevLett.88.027902}, $\ket{\beta}=e^{(\int \beta S(\omega) \hat a^{\dagger}(\omega)-\beta^*S^*(\omega)\hat a(\omega) d\omega)}\ket{0}$. In this case, \eqref{OmegaNSep} becomes $(\Delta \hat \Omega_c)^2 =  |\beta|^2 \int \omega^2 |S(\omega)|^2 d \omega = |\beta|^2\overline {\omega^2}$. This result can be interpreted from different perspectives. In first place, it is proportional to the field's intensity $|\beta|^2$, and corresponds to the shot-noise limit, as expected. In second place, it does {\it not} depend on the total energy of the system $|\beta|^2 \bar \omega$, usually considered as a resource, but rather to the spectral's fluctuations $\overline{\omega^2}$ \cite{Comment}. Thus, for a given fixed field's energy, one can freely engineer its spectrum so as to define different time precision scales using $\Delta \omega$ while keeping the same shot-noise scaling (as done in \cite{Praxmayer,WalmsleySubPlanck, fabre_parameter_2021}, for instance). This suggests that the field's energy should not be considered as the classical ressource, and spectral properties should play a role. We'll study this issue by considering intrinsically multimode states where the frequency variance takes a more complex form. In this case, frequency and intensity properties are not independent and the frequency variance can be used to modify the scaling of the QFI even in a situation where the photon number variance vanishes. 

\begin{figure}
\def\tabularxcolumn#1{m{#1}}

\begin{tabular}{cc}
\subfloat[Separable state, shot-noise scaling]{\includegraphics[width=3.5cm]{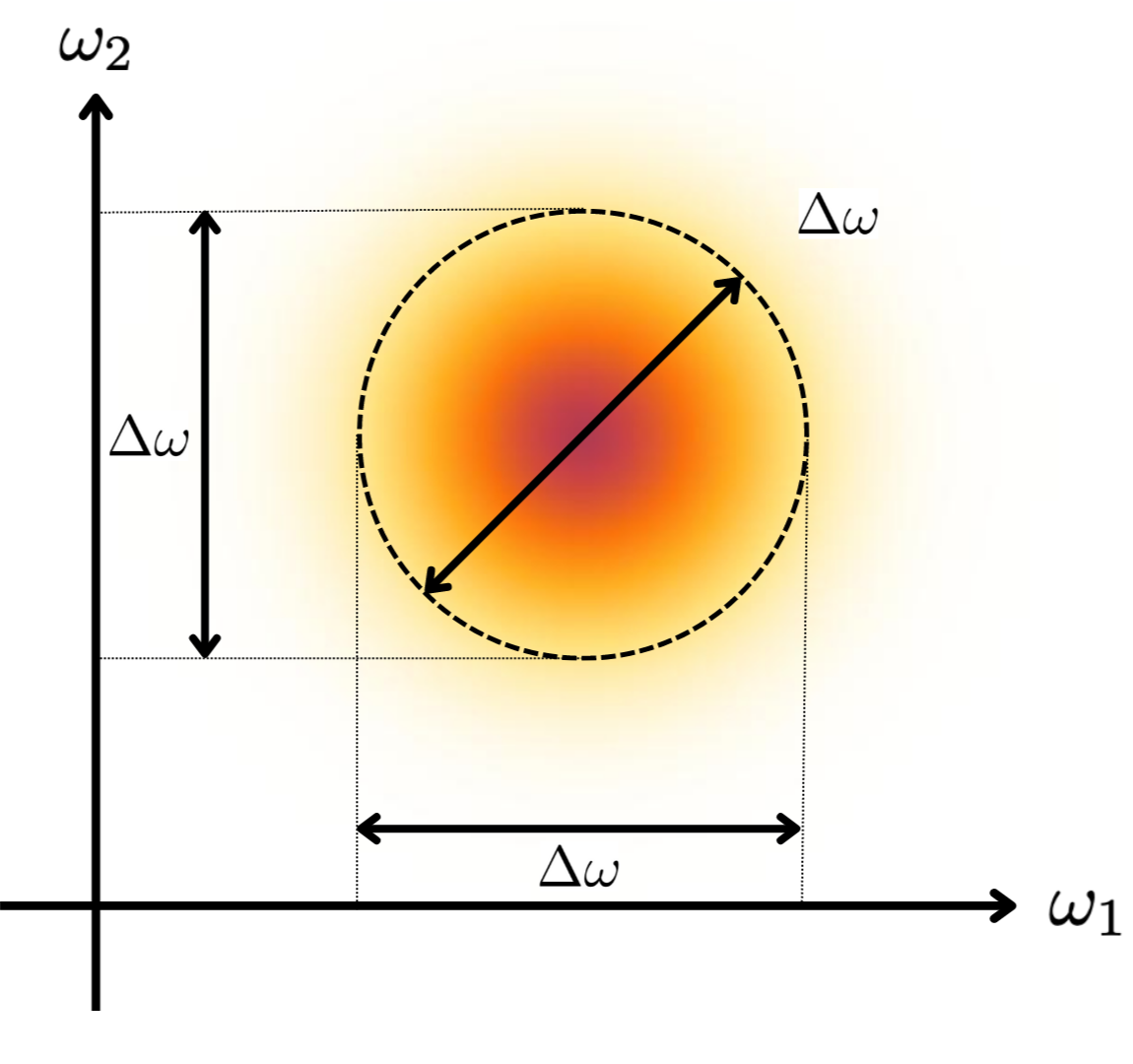}} 
   & \subfloat[Separable state, shot-noise scaling. ]{\includegraphics[width=3.5cm]{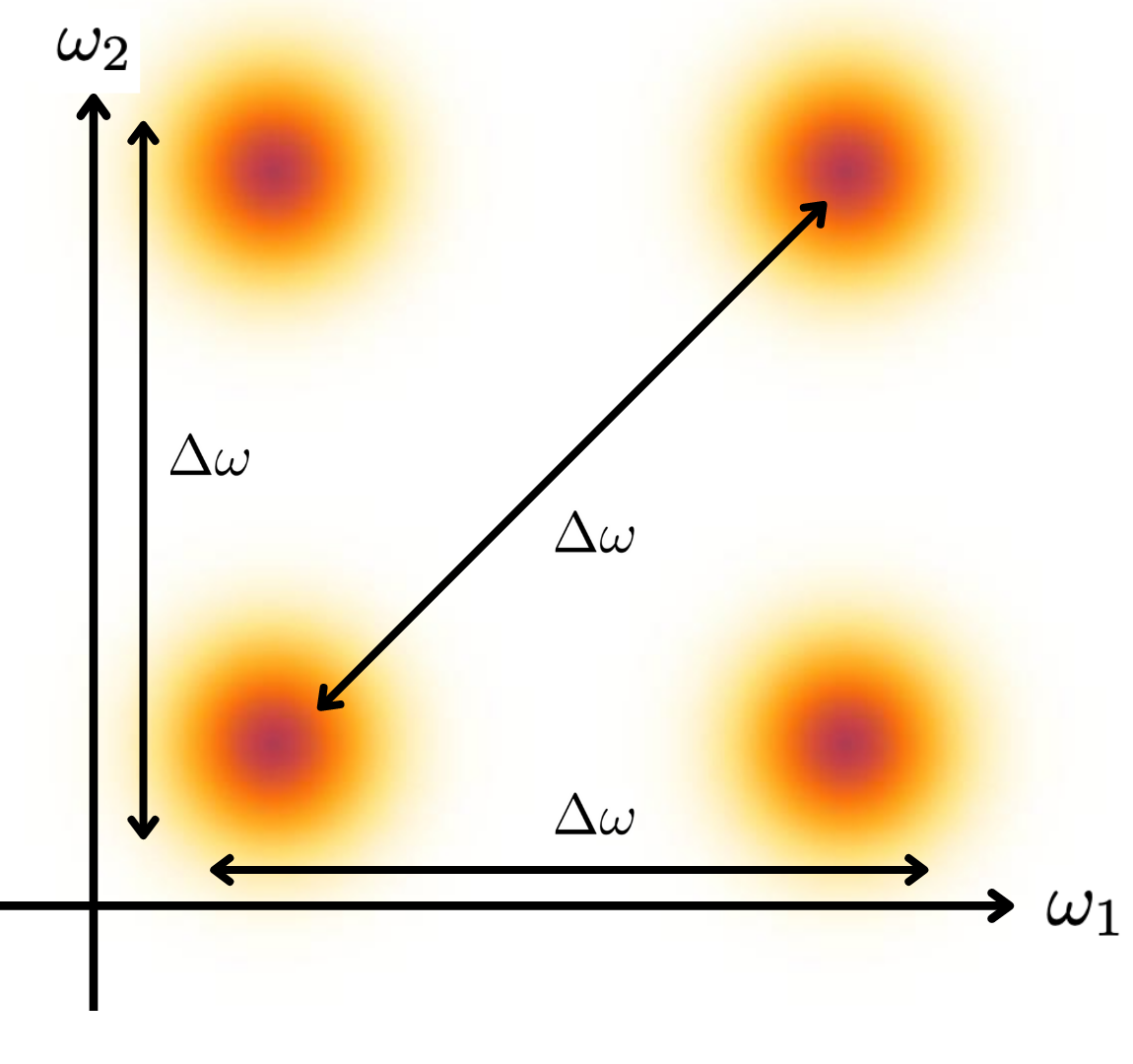}}\\
\subfloat[Entangled state, Heisenberg-like scaling.]{\includegraphics[width=3.5cm]{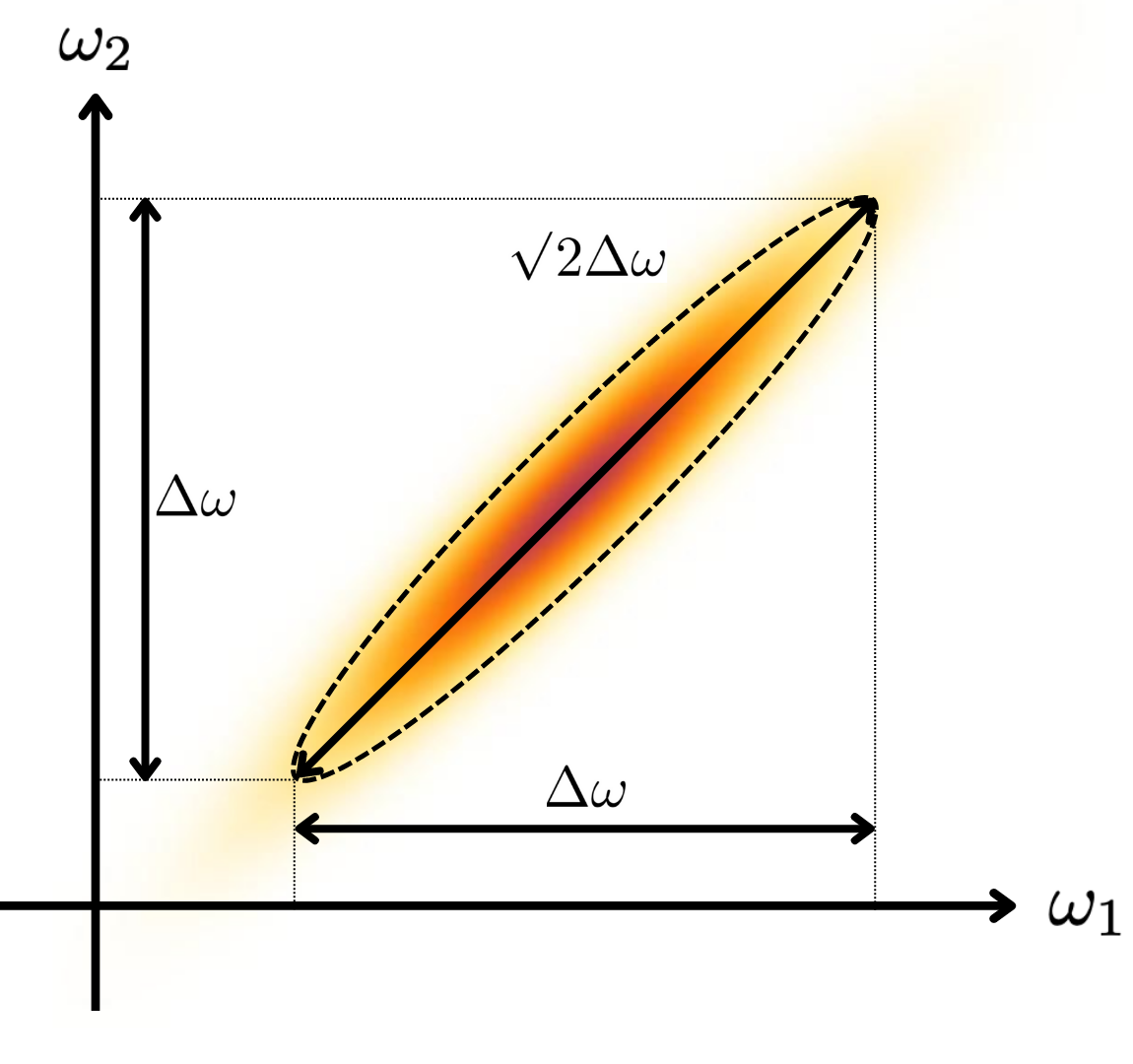}} 
   & \subfloat[Entangled state, shot-noise scaling]{\includegraphics[width=3.5cm]{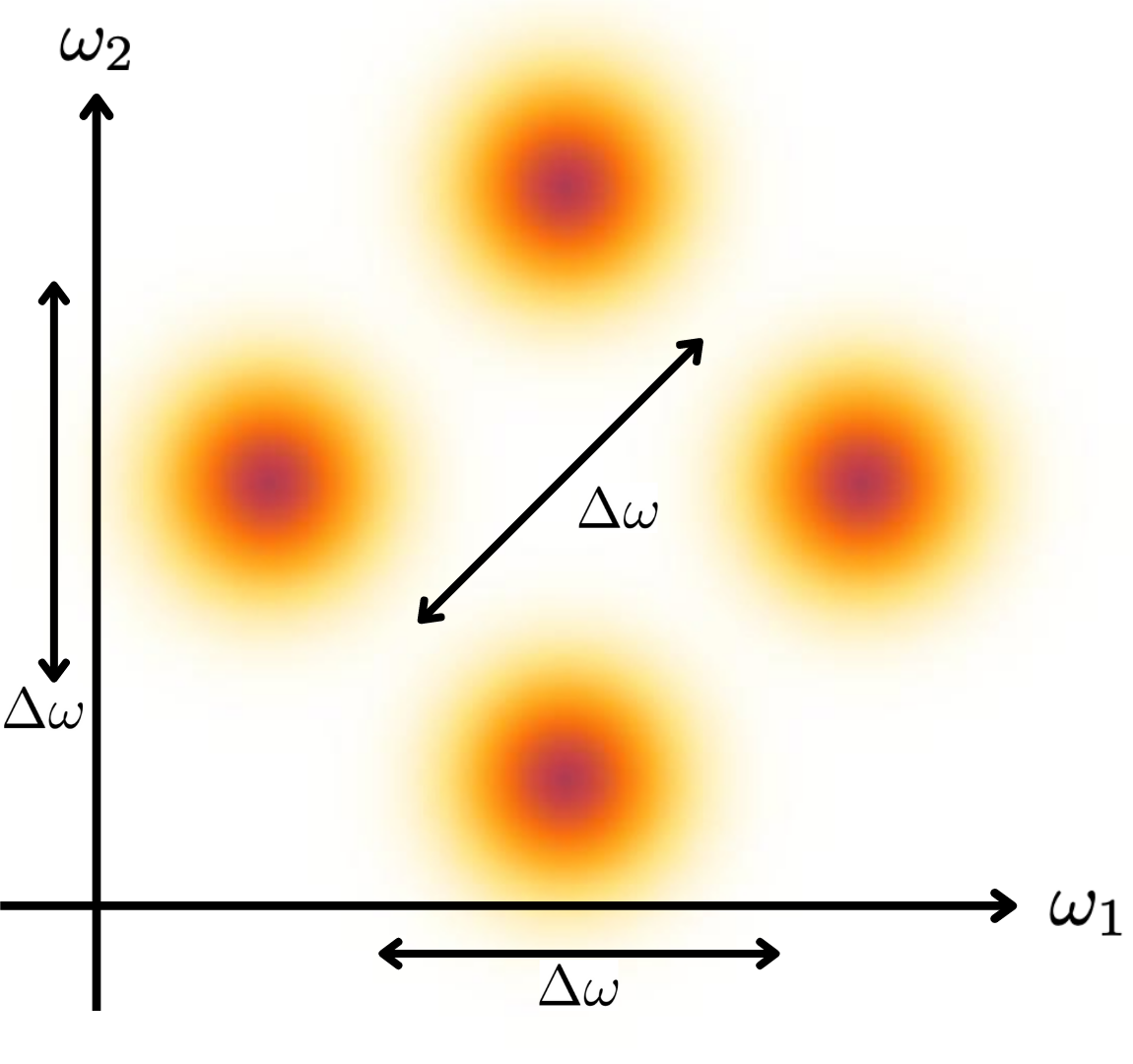}}\\

\end{tabular}

\caption{Joint Spectral Intensity (JSI) of different quantum states ($n=2$). In (a) and (b) we display examples of separable states with a shot-noise scaling and the associated root mean square (RMS). For both, the RMS in the collective variable $\omega_- = \omega_1 - \omega_2$ is equal to $\Delta \omega$. In (c) and (d) entangled states are displayed. In (c) we have a $\sqrt{2}$ scaling factor for a state in the diagonal ($\eta=1$). For $\eta =35/36$ (or $(\Delta \hat \Omega_p)^2 = 12\sigma^2$) we have a physical state with a finite spectral width $\sigma$ in variable $\omega_-$. In (d) the state is entangled but scales as the shot-noise, since the RMS $\Delta \omega_- = \Delta \omega$.}\label{fig1}
\end{figure}

To demonstrate this, we examine a system comprising $n$ single photons, each occupying a distinct ancillary mode. Each photon has a given frequency profile (spectrum), and this system forms a subspace denoted ${\cal S}_n$ (see \cite{PhysRevA.105.052429} and Supplementary Material A). Hence, if photons are prepared in a separable state, $(\Delta \hat \Omega_s)^2 = \sum_{i=1}^n (\Delta \omega_i)^2$, where $(\Delta \omega_i)^2 = \left [\int \omega^2 |S_i(\omega)|^2 d \omega- (\int \omega |S_i(\omega)|^2 d\omega)^2\right ]$, and $S_i(\omega)$ is the spectrum of the $i$-th photon. We'll suppose, for simplicity,  that all the single photons have the same frequency RMS $\Delta \omega$ - also called the frequency RMS {\it per} photon -, and that the considered state is pure (our results can be easily generalized for non-pure states and arbitrary RMS {\it per} photon). Thus,  $(\Delta \hat \Omega_s)^2 = n (\Delta \omega)^2$, since we have the equivalent to $n$ independent probes. This is the same scaling as the shot-noise. By comparing it to the coherent state scaling, we can identify $n (\Delta \omega)^2 = |\beta|^2 \overline {\omega^2}$. Both expressions are proportional to the number of photons: not surprisingly, a coherent state represents the same resource as $n$ independent photons \cite{MacconeNature, PhysRevLett.74.4835, Comment}. Nevertheless, it is noteworthy that while for a coherent state the scaling on the average photon number is due to the fact that $(\Delta \hat n)^2 =|\beta|^2$, $\Delta \hat n = 0$ in ${\cal S}_n$. In Fig. \ref{fig1} (a) and (b) we show the Joint Spectral Intensity (JSI) of separable states of two independent (separable) photons ($n=2$). Also, we can identify the frequency dependency of the coherent state scaling to a frequency variance centered at $\overline \omega = 0$.

We can now calculate the variance of the operator $\hat \Omega$ for a pure non-separable state in ${\cal S}_n$, which gives:
\begin{equation}\label{OmegaN}
(\Delta \hat \Omega)^2 = \sum_{i=1}^n (\Delta \omega_i)^2 +\!\!\!\sum_{i=1, i\neq j}^N \!\!\alpha_i \alpha_j (\langle \hat \omega_i \hat \omega_j \rangle -\langle \hat \omega_i\rangle \langle \hat \omega_j \rangle).
\end{equation}
This variance is bounded, and in the case where all the variances of the single photons are the same we can show that $(\Delta  \hat \Omega)^2 \leq n^2 (\Delta \omega)^2$, which corresponds precisely to the Heisenberg limit. We now compare this result to the usual computations of precision limits in phase measurements in quantum optics, where the role of mode variance is disregarded as a quantum resource and the quantum metrological advantage is exclusively due to the photon number variance: for NOON \cite{Kok_2004, TanzNPJ} or Schr\"odinger cat states \cite{Gilchrist_2004}, which saturate the Heisenberg limit, $(\Delta \hat n)^2 \propto \langle \hat n \rangle ^2$, where $\langle \hat n \rangle$ is the average photon number. For states in ${\cal S}_n$, however, the photon number variance is always equal to zero and the variance in the global evolution generator $\hat \Omega$ explicitly depends on modal properties only. Thus, the Heisenberg limit can only be reached by exploiting mode (frequency) entanglement and the associated mode/particle statistical properties of an intrinsically multi-mode state. Interestingly, in order to reach the Heisenberg limit, these variables must behave as maximally correlated classical ones. However, this is by no means a paradox:  since the considered states are pure and because of the single photon statistics, these correlations effectively contribute to the QFI, leading to the possibility to attain the Heisenberg limit.

We now discuss in detail the type of states that saturate the Heisenberg limit for \eqref{OmegaN} and provide a geometrical intuitive picture of the observed scaling, pointing out the analogies and differences with respect to the quadrature phase space \cite{zurek_sub-planck_2001, LuizSub}. These states, also discussed in \cite{MacconeNature}, are maximally entangled in the (local) variables $\omega_i$, and their general mathematical expression (see Supplementary Material B) for $\alpha_i =1 ~\forall ~i$ reads:
\beq\label{states}
\ket{\psi}=\int d\Omega f(\Omega) \ket{\Omega + \omega_1^0}\ket{\Omega + \omega_2^0}...\ket{\Omega + \omega_n^0},
\eeq
where $\omega_i^0$ are constants. The spectral function thus only depends on one variable ($\Omega$), and $\ket{\psi}$ has a (non-physical) spectrum that is infinitely localized in all collective variables except for $\Omega = \sum_{i=1}^n \omega_i$, the one associated to the operator $\hat \Omega$.  This means that all the photons display a collective behavior associated to a re-scaled de Broglie wavelength $\lambda = c / \Omega$ \cite{PhysRevLett.74.4835}. As can be seen from the JSI shown in Fig. \ref{fig1}(c) (for $n=2$), these states are represented by diagonals, and the variance of each mode $i$ is the projection of these diagonals on the corresponding frequency axis. This geometrically illustrates the role of correlations in the scaling. This type of states with different spectral functions is currently produced in experiments for $n=2$ (see \cite{chen_hong-ou-mandel_2019,PhysRevLett.103.253601,Qudits, PhysRevA.102.012607, PhysRevA.82.013804}, for instance and Sup. Mat. sections D and E). In addition, from Fig. \ref{fig1} (c) and (d) we see that entanglement, even if necessary, is not sufficient to obtain sub-shot-noise scaling, and that the symmetry of the spectral variance plays an important role on the state's metrological precision. In the case where we have $n$ photons in $n$ modes, the same type of geometrical picture can be built, and the states saturating the Heisenberg limit are diagonals of a $n$ dimensional hypercube. 

These scaling effects can also be observed in the time frequency phase space (TFPS). For this, we define the Wigner function of a general  state in ${\cal S}_n$ as \cite{Brecht:13, douce_direct_2013, fabre_parameter_2021}: 
\begin{eqnarray}\label{Wigner}
&&W(\phi_1,...,\phi_n, \tau_1,...,\tau_n)= \int d\omega_1 ... \int d\omega_n e^{2i\sum_{i=1}^n \omega_i\tau_i}\times \nonumber\\
&&\bra{\phi_1+\omega_1,...\phi_n+\omega_n}\hat \rho \ket{\phi_1-\omega_1,...,\phi_n-\omega_n},
\end{eqnarray}
where $\hat \rho$ is the $n$ photon state in ${\cal S}_n$ and $\omega_i$ refers to the frequency variable of the photon occupying the $i$-th spatial mode. 

From expression \eqref{Wigner}, we see that the operator $\hat \Omega$ implements {\it translations} in TPFS, {\it i.e.}, $\tau_i \rightarrow \tau_i+\alpha_i \delta t$, where $\delta t$ is the parameter to be estimated. By inspecting Eq. \eqref{OmegaNSep}, and considering the mapping between separable states in ${\cal S}_n$ and coherent states, we can see that while for a coherent monochromatic field the precision limit is set by its {\it rotation} in the quadrature phase space, for a non-monochromatic single-mode one it is set by a combination of a rotation and a {\it translation} in TFPS. However, separable states in ${\cal S}_n$ ($n$ independent single photon states) have a rotational symmetry in the quadrature phase space, so their metrological power cannot come from this space: it is exclusively associated to their translations in the TFPS. Finally, for states saturating the Heisenberg limit as \eqref{states}, since they can be described by a single wave-function in variable $\Omega$, their corresponding Wigner function (defined by \eqref{Wigner}) becomes (see Supplementary Material C): $ W(\phi_1+\omega_1^0,...,\phi_n+\omega_n^0,\tau_1,...,\tau_n) = W_1(\phi_1,\tau_1+\tau_2+...\tau_n)\times \delta(\phi_2-\phi_1)...\delta(\phi_n-\phi_{n-1})$ and can be entirely described in a two dimensional phase space. Nevertheless, due to the collective nature of the associated variables, the translation of each $\tau_i$ by the same amount $\delta t$ ($\alpha_i = 1$) results in a translation of the Wigner function by an amount $n\delta t$. This behavior, which can be seen as a change of scale in the temporal variable, is a consequence of the correlations between photons, and is the TFPS signature of the Heisenberg scaling of the QFI. Thus, with a displacement of $\delta t/n$ in the TFPS associated to the collective variable $\Omega$, the Wigner function assumes its value on $\tau + \delta t$, {\it i.e.}, and evolves $n$ times ``faster" than for translations on independent variables, providing a TFPS picture of the metrological quantum advantage that is a multi-dimensional analogous to the one introduced in \cite{LuizSub, zurek_sub-planck_2001} but comes from an entirely different physical property. The TFPS representation of quantum metrological aspects of the field can only be observed in multi-mode states and appear when describing the field's spectral properties using collective variables. Consequently, in TFPS, sub-Planck-like structures \cite{Praxmayer,WalmsleySubPlanck} cannot be associated to any quantum effect, even if they're interesting to optimize the variance and consequently improve the metrological applications of classical and single-mode fields given a certain spectral bandwidth.  Notice that it is also possible to use collective variables and a two dimensional phase space to describe the shot-noise scaling for independent photons using the same techniques described above, but in this case, translations implemented by the collective operator $\hat \Omega$ translate the Wigner function by an amount $\sqrt{n}\delta t$ only. 

The states reaching the Heisenberg limit are non-physical, since they're maximally correlated and represented by Dirac distributions. Physical states must have a finite spectral width, and to see how this affects our results we'll consider for simplicity that the spectrum of all the other $n-1$ collective variables (different from the one associated to the operator $\hat \Omega$) are distributions with a same variance, $\sigma^2$. In this case, it can be shown (see Supplementary Material B), that $(\Delta \hat \Omega_p)^2 = n^2((\Delta \omega)^2 - \sigma^2)+n\sigma^2$ and the Heisenberg limit is no longer reached. As a matter of fact, this expression displays a transition between a quadratic to a linear behavior in $n$. This can be seen by defining $\eta \in \left[0,1\right]$ - where $\eta = 1$ corresponds to the ideal case of non-physical states, and $\eta = 0$ to the non-correlated (separable) state -, and $\sigma^2 = (1-\eta)( \Delta \Omega_p)^2/n$. Hence, $(\Delta \Omega_p)^2= n^2 (\Delta \omega)^2 /(n(1-\eta)+\eta)$ and a quadratic to linear transition occurs for $n \approx \eta/(1-\eta)$: the variance follows mostly a quadratic behavior for $n \ll \eta/(1-\eta) $ and a linear one for $n \gg  \eta/(1-\eta)$. We can have an idea of this effect by considering that $\eta = 0.99$ (see Supplementary Material E), a limit  that can be reached with no difficulty for $n=2$ in many experimental set-ups \cite{Qudits, Orieux_2017, chen_hong-ou-mandel_2019,jeannic_experimental_2021}. In this case, a predominant quadratic scaling is ensured for $n \lesssim 99$, showing that sub-shot-noise scaling is quite robust when realistic states are considered. A more complete discussion on the values of $\eta$ in experimental set-ups can be found in Supplementary Material D and E.   

The existence of a transition from the Heisenberg scaling to the shot-noise one recalls the results obtained in  \cite{LuizMetro} in a completely different context, where the authors considered a photon loss model which was also controlled by a parameter $\eta$. In that case, $\eta=0$ represented the situation of maximal loss while $\eta = 1$  the situation of no loss. Here, we're dealing with pure states only, but the finite spectrum of continuous variables can be seen as a continuous superposition with some width of frequency displaced states. In quantum computing and quantum information models with continuous variables, such displacements, and consequently the finite width of distributions, are considered as errors/deviations from the ideal case \cite{gottesman_encoding_2001,PhysRevLett.112.120504,PhysRevA.102.012607}. They produce effects similar to the ones caused by losses, even though the considered states are pure. This fact suggests a beautiful connection between {\it physical} continuous variables states ({\it i.e.}, states with spectral distributions of finite width), noise models for continuous variables quantum information and ultimate precision limits on noisy quantum metrology, and will be the subject of future work. 

As a conclusion, we have presented a physical model clearly exhibiting the subtleties of the interplay between modes and the quantum field's statistics. We discussed this issue in the framework of metrology and we have shown that a quadratic scaling with the number of resources can appear in situations where the field is multimode and entangled but has zero photon number variance. We have done so by studying frequency correlations in single photon states, so the number of modes is directly associated to the number of photons and mode correlations inherit from the photons' quantum nature - in the same way as polarization correlations, when associated to single photons, display non-classical features \cite{aspect_experimental_1982}.  We then discussed a geometrical interpretation of our results, distinguishing two types of contributions to the scaling of the time precision limits: one coming from a collective but independent effect, and another from a collective quantum effect that leads to a (effective) transition of the $n$ variable system to a single variable one. Finally, we discussed the effects of a finite spectrum and suggested a relation between spectral width and noise. It's important to notice that in the present work we ruled out any quantum advantage coming from spectral properties of single mode fields, but rather associated them to a classical resource (a coherent state). We have focused on time estimation and frequency variables, but our results can be generalized to other pairs of continuous degrees of freedom of single photons, as the transverse position and momentum \cite{tasca_continuous_2011}. An interesting and challenging perspective is to generalize our techniques to study the metrological properties of more complex multimode non-Gaussian states and investigate the appearance of a mode dependent particle number-like dependent scaling.

\section*{Acknowledgements}

We acknowledge funding from the Plan France 2030 through the project ANR-22-PETQ-0006.

\onecolumngrid
\appendix
\section{A. Basics}

A single photon pure state at mode $i$ with frequency $\omega$ is described by the application of the creation operator to the vacuum state: $\hat{a}_{i}^{\dagger}(\omega)\ket{\text{vac}}=\ket{\omega}_i$. The label $i$ can be polarization, a spatial mode - as the transverse propagation direction -, or any other combination of modes that plays the role of an ancillary mode that creates distinguishability between each photon.  We can  also define the annihilation operator such that $\hat{a}_i(\omega)\ket{\omega'}_i=\delta(\omega-\omega')\ket{\text{vac}}$. In addition, the commutation relation between creation and annihilation operator is given by:
 
 \begin{equation}\label{commutationrl}
 [\hat{a}_{\alpha}(\omega),\hat{a}_{\beta}^{\dagger}(\omega')]=\delta(\omega-\omega')\delta_{\alpha\beta}\mathds{I},
 \end{equation}
 where $\alpha$ and $\beta$ are auxiliary modes. We also have that   $[\hat{a}_{\alpha}(\omega),\hat{a}_{\beta}(\omega')]=0$ and $[\hat{a}^{\dagger}_{\alpha}(\omega),\hat{a}^{\dagger}_{\beta}(\omega')]=0$.

If we consider to be in the narrowband approximation \cite{PhysRevA.72.032110}, so that the central frequency of the spectral distribution is much larger than its spectral width, integrals can be extended over the whole frequency spectrum, and the Fourier transform of the annihilation operator is the annihilation operator at the arrival time $t$:

 \begin{equation}\label{Fourier}
 \hat{a}(t)=\frac{1}{\sqrt{2 \pi}}\int_{\mathds{R}} d\omega \hat{a}(\omega) e^{-i\omega t},
 \end{equation}
 the same being valid for the creation operation, of course. We also have that 
  \begin{equation}\label{commutationrltime}
 [\hat{a}_{\alpha}(t),\hat{a}_{\beta}^{\dagger}(t')]=\delta(t-t')\delta_{\alpha\beta}\mathds{I},
 \end{equation}
 where $\alpha$ and $\beta$ are auxiliary modes, and   $[\hat{a}_{\alpha}(t),\hat{a}_{\beta}(t')]=0$ and $[\hat{a}^{\dagger}_{\alpha}(t),\hat{a}^{\dagger}_{\beta}(t')]=0$.

 We stress that in the present description time is seen not as a parameter but as a degree of freedom associated to the  arrival time of photons in a detector.

 A general single photon pure state can be decomposed in the time  basis or, equivalently, in the spectral basis as,
\begin{equation}\label{state}
\ket{\psi}=\int_{\mathds{R}} d\omega S(\omega) \hat{a}^{\dagger}(\omega)\ket{\text{vac}}.
\end{equation}
The amplitude spectrum $S(\omega)$ is the Fourier transform of the time of arrival distribution and $\abs{S(\omega)}^{2}=\abs{\bra{\omega}\ket{\psi}}^{2}$ denotes the probability density of detecting a photon with frequency $\omega$.  We can, of course, also construct from this principles  general mixed single photon states described by a density matrix.

The space of states we consider in the present contribution consists of a collection of $n$ single photon states in $n$ different ancillary modes. This space will be called ${\cal S}_n$, where $n$ is the number of distinguishable modes and also the number of photons. It means that only cases where there is at most one photon per mode are considered. 

A general pure state in ${\cal S}_n$  can be written as
\begin{equation}\label{State}
\ket{\psi}=\int d\omega_{1}...\int d\omega_{n}  F(\omega_{1},...,\omega_{n}) \hat{a}_{1}^{\dagger}(\omega_{1})...\hat{a}_{n}^{\dagger}(\omega_{n})\ket{0}.
\end{equation}
where the spectral function $F(\omega_{1},...,\omega_{n})$ is normalized to one : $\int d\omega_{1}...\int d\omega_{n} | F(\omega_{1},...,\omega_{n})|^2 =1$. 

The time and frequency operators are defined as:

\begin{equation}
\hat{t}_{a}=\int_{\mathds{R}} t \hat{a}^{\dagger}(t)\hat{a}(t) dt,\\
\hat{\omega}_{a}=\int_{\mathds{R}} \omega \hat{a}^{\dagger}(\omega)\hat{a}(\omega) d\omega.
\end{equation}
When applied to single photons states, these operators fulfill the eigenvevtors-eigenvalues equation : $\hat{t}_{a}\ket{t}_{a}=t\ket{t}_{a}$ and $\hat{\omega}_{a}\ket{\omega}_{a}=\omega \ket{\omega}_{a}$. The frequency operator is proportional to the  free Hamiltonian $\hat{E}_{a}=\hbar \hat{\omega}_{a}$. 
As previously, we considered the narrowband approximation of a photon with central frequency far from origin. Consequently, the integration over the frequency can safely be considered as covering all $\mathds{R}$. 
As for the time variable, it corresponds to the Fourier transform of frequency for all practical purposes and is physically associated to the time of detection conditioned to the fact that a detection has indeed happened \cite{giovannetti_quantum_2015,maccone_quantum_2020}. 

Using Eq.~(\ref{commutationrl}), we can see that time and frequency operators do not commute in the single photon  (single mode) regime:

 \begin{equation}\label{noncommutation}
 [\hat{\omega}_{a},\hat{t}_{a}]=i\mathds{I}. 
 \end{equation}
They form, together with the identity operator $\mathds{I}$, a three-dimensional Heisenberg algebra in perfect analogy with the position and momentum operators. This fact is not true in general for modes occupied by more than one photon and it is essential for building a set of universal gates which manipulate frequency and time as the universal gates defined for position and momentum manipulate states defined in these basis \cite{PhysRevA.105.052429}.

 \section{B. From non-physical states saturating the Heisenberg limit to physical states}
 
 A state saturating the Heisenberg limit can be obtained by noting that the majoration $(\Delta  \hat \Omega)^2 \leq n^2 (\Delta \omega)^2$ can be obtained by the Cauchy-Schwarz (CS) inequality, applied to the covariance $\operatorname{Cov}(\omega_i,\omega_j)=\langle \hat \omega_i \hat \omega_j \rangle -\langle \hat \omega_i\rangle \langle \hat \omega_j \rangle$. We thus know from the case of equality in the CS bound that for all $i$ and $j$ there must exist a constant $\lambda_{i,j}$ such that $\Delta(\omega_i-\lambda_{i,j}\omega_j)=0$, meaning that if we treat the $\omega_i$ as random variables, $\omega_i-\lambda_{i,j}\omega_j$ must be constant. Since we assume that $\operatorname{Cov}(\omega_i,\omega_j)=(\Delta\omega)^2$, we have $\lambda_{i,j}=1$.\\
These relations between the $\omega_i$'s impose that the JSA of the state is a product of many $\delta$ functions. All these relations allowing only one remaining degree of freedom, we get the general expression of the state achieving the Heisenberg bound:
\begin{equation}
    \ket{\psi}=\int d\omega_1...d\omega_n f(\omega_1+...+\omega_n)\delta(\omega_1-\omega_2+C_1)...\delta(\omega_{n-1}-\omega_n+C_{n-1})\ket{\omega_1}\ket{\omega_2}...\ket{\omega_n}.
\end{equation}
This state can be expressed in terms of the collective variable $\Omega=\omega_1+...+\omega_n$, and integrating the delta functions, we get the general formula:
\begin{equation}\label{inf}
    \ket{\psi}=\int d\Omega f(\Omega) \ket{\Omega + \omega_1^0}\ket{\Omega + \omega_2^0}...\ket{\Omega + \omega_n^0}.
\end{equation}

Notice that state \eqref{inf} corresponds to choosing $\alpha_i =1$$\forall i$, but different collective variables (with different distributions for the coefficients $\alpha_i=\pm 1$) are possible by making a proper choice of the constants $C_i$. 

As for maximally correlated non-pure states, even though they have the same variance as maximally entangled ones, the QFI for mixed states {\it is not} related to the variance. Thus, we can show that the scaling of the associated QFI for these states is the same as the one for separable states, saturating the bounds found in \cite{PhysRevA.87.032324,Bound2}.

Of course, since \eqref{inf} is  infinitely concentrated in the collective variable  $\Omega=\omega_1+...+\omega_n$, it is not physical. We can turn it into a physical state by considering that the width $\sigma$ in variables other that $\Omega$ are different from zero. This correspond to replacing the delta functions by (for example) Gaussian functions.
 To simplify the computation, we consider an orthonormal basis $p_i$, with $p_1=\frac{1}{\sqrt{n}}(\omega_1+...+\omega_n)$ being the collective variable associated to the generator of the evolution and the other $p_{i}$, $i\neq 1$ complete the orthonormal basis. We then look at the state:
\begin{equation}
    \ket{\psi}=\int d\omega_1 ... d\omega_n f(p_1)g(p_2)\cdots g(p_n)\ket{\omega_1,...,\omega_n}
\end{equation}
with the functions $\abs{f}^2$ and $\abs{g}^2$ having respectively a width of $\Delta$ and $\sigma$, with the assumption that $\Delta \gg \sigma$. So, when computing the expectation values, we have: $(\Delta p_1)^2=\Delta^2$, $(\Delta p_i)^2=\sigma^2$ for $i>1$ and ${\rm Cov}(p_i,p_j)=0$.\\

We can then compute the variance of $\omega_i$ and $\Omega=\omega_1+...+\omega_n=\sqrt{n}p_1$. The last one is the simplest one:
\begin{equation}
    (\Delta\hat\Omega)^2=n(\Delta p_1)^2=n\Delta^2
\end{equation}
The variance of $\omega_i$ is a little bit more lengthy. To simplify the computation, we view the variable $\omega_i$ and $p_j$ as vectors of an $n$ dimensional vector space with the canonical basis $\{\omega_i\}$. This allow us to use the scalar product notation to simplify the expression of one type of variables in term of the other. More specifically we write:
\begin{equation}
    \omega_i=(\omega_i\mid p_1) p_1+...+(\omega_i\mid p_n) p_n
\end{equation}
So:
\begin{subequations}
    \begin{align}
        (\Delta\omega_i)^2&=\Delta\big[(\omega_i\mid p_1) p_1+...+(\omega_i\mid p_n) p_n\big]\\
        &= (\omega_i\mid p_1)^2\underbrace{(\Delta p_1)^2}_{\Delta^2}+(\omega_i\mid p_2)^2\underbrace{(\Delta p_2)^2}_{\sigma^2}+\cdots+(\omega_i\mid p_n^2)^2 \underbrace{(\Delta p_n)^2}_{\sigma^2}\\
        &=\underbrace{(\omega_i\mid p_1)^2}_{1/n}(\Delta^2-\sigma^2)+\sigma^2\underbrace{\sum_j (\omega_i\mid p_j)^2}_{1}\\
        &=\frac{1}{n}(\Delta^2-\sigma^2)+\sigma^2
    \end{align}
\end{subequations}
In the case where $\Delta \omega_i = \Delta \omega \: \forall \: i$, we obtain $(\Delta \hat \Omega_p)^2 = n^2((\Delta \omega)^2 - \sigma^2)+n\sigma^2$, where the subscript $p$ stands for ``physical state". By setting $\sigma^2=(1-\eta)(\Delta\hat\Omega_p)^2/n=(1-\eta)\Delta^2$, we have:
\begin{equation}
   ( \Delta\hat\Omega_p)^2=\frac{n^2}{n(1-\eta)+\eta}(\Delta\omega)^2
\end{equation}

\begin{figure}
\includegraphics[width=8cm]{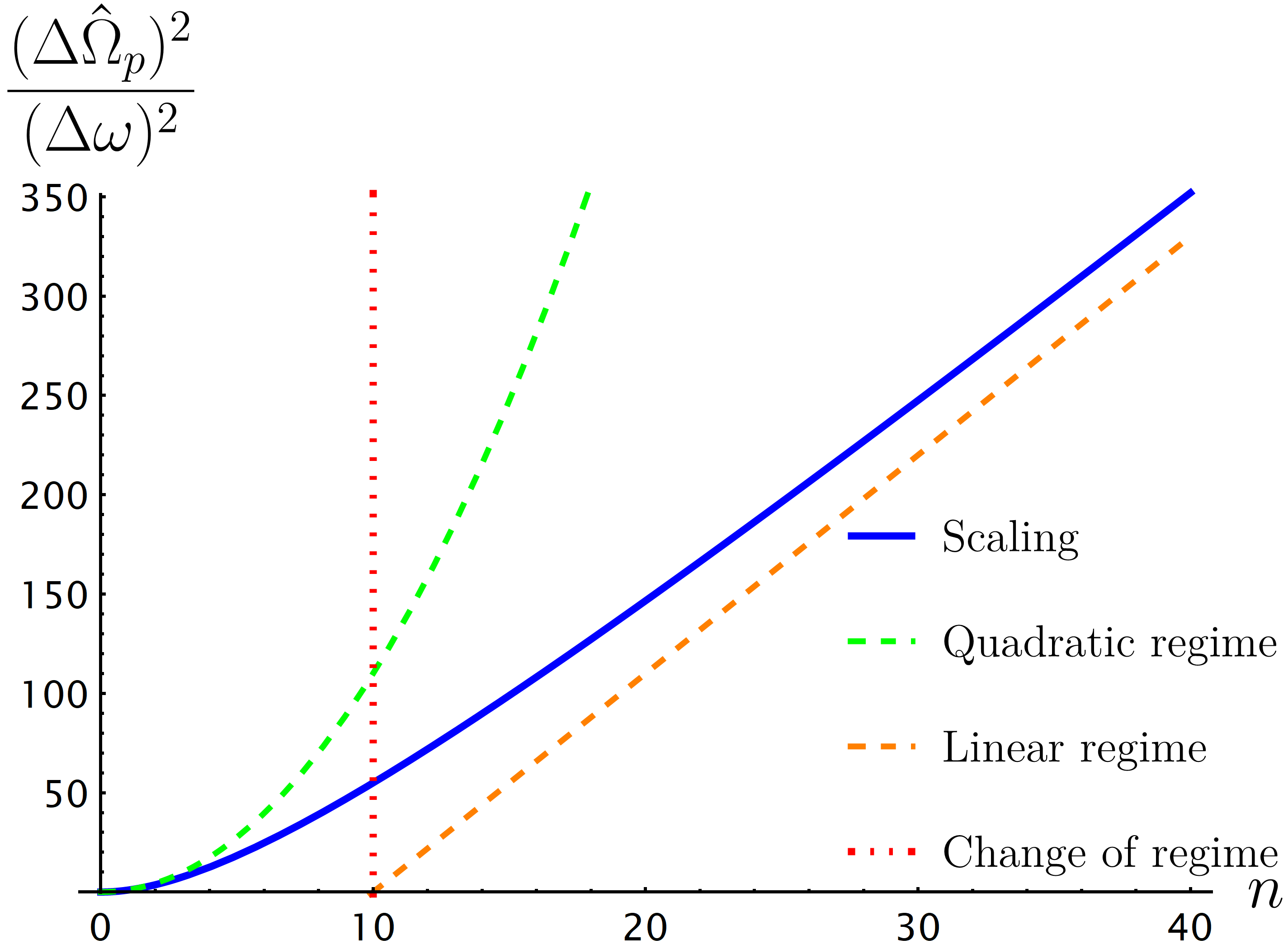}
\caption{$(\Delta \hat \Omega_p)^2/(\Delta \omega)^2$ as a function of the number of photons. We display the different types of scalings (linear and quadratic) as well as the variance and the transition point, the value of $n$ for which one changes from the quadratic to the linear scaling. }
\label{figScaling}
\end{figure}

We see that the gain is quadratic in the case $\eta=1$ ($\sigma=0$) and linear for $\eta=0$. For an intermediate value of $\eta$, the scaling is quadratic for small $n$, and then becomes linear for $n$ large enough. The transition happens when $\frac{1}{n^2}\left(1-\frac{\sigma^2}{\Delta^2}\right)=\frac{1}{n}\frac{\sigma^2}{\Delta^2}$ {\it i.e.} when $n=\frac{\eta}{1-\eta}$. We see in Fig. \ref{figScaling} the transition between the two regimes as a function of $n$ for $\eta \approx 0.91$. For $\eta=0.99$, the transition from the Heisenberg limit occurs for $n\approx 100$ photons as mentioned in the main text.

\section{C. The Wigner function representation of scaling}
 
For the diagonal state:
\begin{equation}
    \ket{\psi}=\int d\Omega f(\Omega)\ket{\Omega+\omega_1^0}...\ket{\Omega+\omega_n^0},
\end{equation}
the Wigner function can be computed as:
\begin{subequations}
    \begin{align}
        W(\phi_1,&...,\phi_n,\tau_1,...,\tau_n)\\
        &=\int d\omega_1...d\omega_n e^{2i(\omega_1\tau_1+...+\omega_n\tau_n)}\bra{\phi_1+\omega_1,...,\phi_n+\omega_n}\ket{\psi}\bra{\psi}\ket{\phi_1-\omega_1,...,\phi_n-\omega_n}\\
        &=\int d\omega_1...d\omega_n d\Omega d\Omega' e^{2i(\omega_1\tau_1+...+\omega_n\tau_n)}f(\Omega)f^\ast(\Omega')\notag\\
        &\braket{\phi_1+\omega_1}{\Omega+\omega_1^0}...\braket{\phi_n+\omega_n}{\Omega+\omega_n^0}\braket{\Omega'+\omega_1^0}{\phi_1-\omega_1}...\braket{\Omega'+\omega_n^0}{\phi_n-\omega_n}\\
        &=\int d\omega_1...d\omega_n d\Omega d\Omega' e^{2i(\omega_1\tau_1+...+\omega_n\tau_n)}f(\Omega)f^\ast(\Omega')\notag\\
        &\delta(\phi_1+\omega_1-\Omega-\omega_1^0)...\delta(\phi_n+\omega_n-\Omega-\omega_n^0)\delta(\Omega'+\omega_1^0-\phi_1+\omega_1)...\delta(\Omega'+\omega_n^0-\phi_n+\omega_n)\\
        &=\int d\Omega d\Omega'\exp[2i(\phi_1-\Omega'-\omega_1^0)\tau_1+...+2i(\phi_n-\Omega'-\omega_n^0)\tau_n]f(\Omega)f^\ast(\Omega')\notag\\
        &\delta(2\phi_1-2\omega_1^0-\Omega-\Omega')...\delta(2\phi_n-2\omega_n^0-\Omega-\Omega')\\
        &=\int d\Omega \exp[2i(-\phi_1+\omega_1^0+\Omega)\tau_1+...+2i(\phi_n-2\phi_1+2\omega_1^0+\Omega-\omega_n^0)\tau_n]f(\Omega)f^\ast(2\phi_1-2\omega^0_1-\Omega)\notag\\
        &\delta(2\phi_2-2\omega_2^0-2\phi_1+2\omega_1^0)...\delta(2\phi_n-2\omega_n^0-2\phi_1+2\omega_1^0)\\
        &=\frac{1}{2^{n-1}}\int d\Omega \exp[2i(-\phi_1+\omega_1^0+\Omega)\tau_1+...+2i(\phi_1-2\phi_1+2\omega_1^0+\Omega-\omega_n^0)\tau_n]f(\Omega)f^\ast(2\phi_1-2\omega^0_1-\Omega)\notag\\
        &\delta(\phi_2-\omega_2^0-(\phi_1-\omega_1^0))...\delta(\phi_n-\omega_n^0-(\phi_1-\omega_1^0))\\
        &=\frac{1}{2^{n-1}}\int d\Omega \exp[2i(-\phi_1+\omega_1^0+\Omega)\tau_1+...+2i(-\phi_n+\omega_n^0+\Omega)\tau_n]f(\Omega)f^\ast(2\phi_1-2\omega^0_1-\Omega)\notag\\
        &\delta(\phi_2-\omega_2^0-(\phi_1-\omega_1^0))...\delta(\phi_n-\omega_n^0-(\phi_1-\omega_1^0))\\
        &=\frac{1}{2^{n-1}}\int d\Omega e^{2i\Omega(\tau_1+...+\tau_n)}f(\phi_1-\omega_1^0+\Omega)f^\ast(\phi_1-\omega^0_1-\Omega)\delta(\phi_2-\omega_2^0-(\phi_1-\omega_1^0))...\delta(\phi_n-\omega_n^0-(\phi_1-\omega_1^0)).
    \end{align}
\end{subequations}
Which can also be written as
\begin{equation}
    W(\phi_1+\omega_1^0,...,\phi_n+\omega_n^0,\tau_1,...,\tau_n)=\frac{1}{2^{n-1}}W_1(\phi_1,\tau_1+\tau_2 +\cdots + \tau_n)\delta(\phi_2-\phi_1)...\delta(\phi_n-\phi_1),
\end{equation}
where
\begin{equation} \label{Wig}
W_1(\phi,t) = \int d\Omega e^{2i\Omega t}f(\phi_1+\Omega)f^\ast(\phi_1-\Omega)
\end{equation}
is the Wigner function associated to the spatial single mode $\int d\omega f(\omega)\ket{\omega}$.

If we want to represent the Wigner function $W$  in different directions, the delta functions will render this task impossible. Moreover, the information brought by the delta functions is nothing but the representation of well-defined frequencies, and the associated phase space structure that brings no further information than this one. Since the state is separable in the chosen variables, so is the Wigner function, and we can concentrate our discussion on the Wigner function associated to the only variable that can provide an interesting picture of the state, which is $W_1(\phi_1,t)$ but calculated at $t = \tau_1+...+\tau_n$ (see \eqref{Wig}).

If in addition, since we're looking at an evolution associated to a collective variable, we have  $\tau=\tau_1=\tau_2=\cdots=\tau_n$, so $t=n\tau$, and the relevant Wigner function to look at is:
\begin{equation}\label{WD}
    W_d(\phi_1,\tau)=W_1(\phi_1,n\tau)=\int d\Omega e^{2in\Omega\tau}f(\phi_1+\Omega)f^\ast(\phi_1-\Omega).
\end{equation}
Notice that this situation is different from the one where we only consider the evolution generated by the evolution operator associated to only one variable, which is associated to the term $W_1(\phi_1 - \omega_1^0, \tau)$ of the total Wigner function.
The obtained relation $W_d(\phi,\tau)=W_1(\phi,n\tau)$ \eqref{WD} means that the Wigner function in the collective variable is re-scaled in the $\tau$ direction by a factor $n$ and consequently, that a frequency measurement is indeed more efficient. 

\begin{figure}
\def\tabularxcolumn#1{m{#1}}

\begin{tabular}{ccc}
\subfloat[Wigner function in variable $\omega_1$ for a Schr\"odinger cat-like state]{\includegraphics[width=4.5cm]{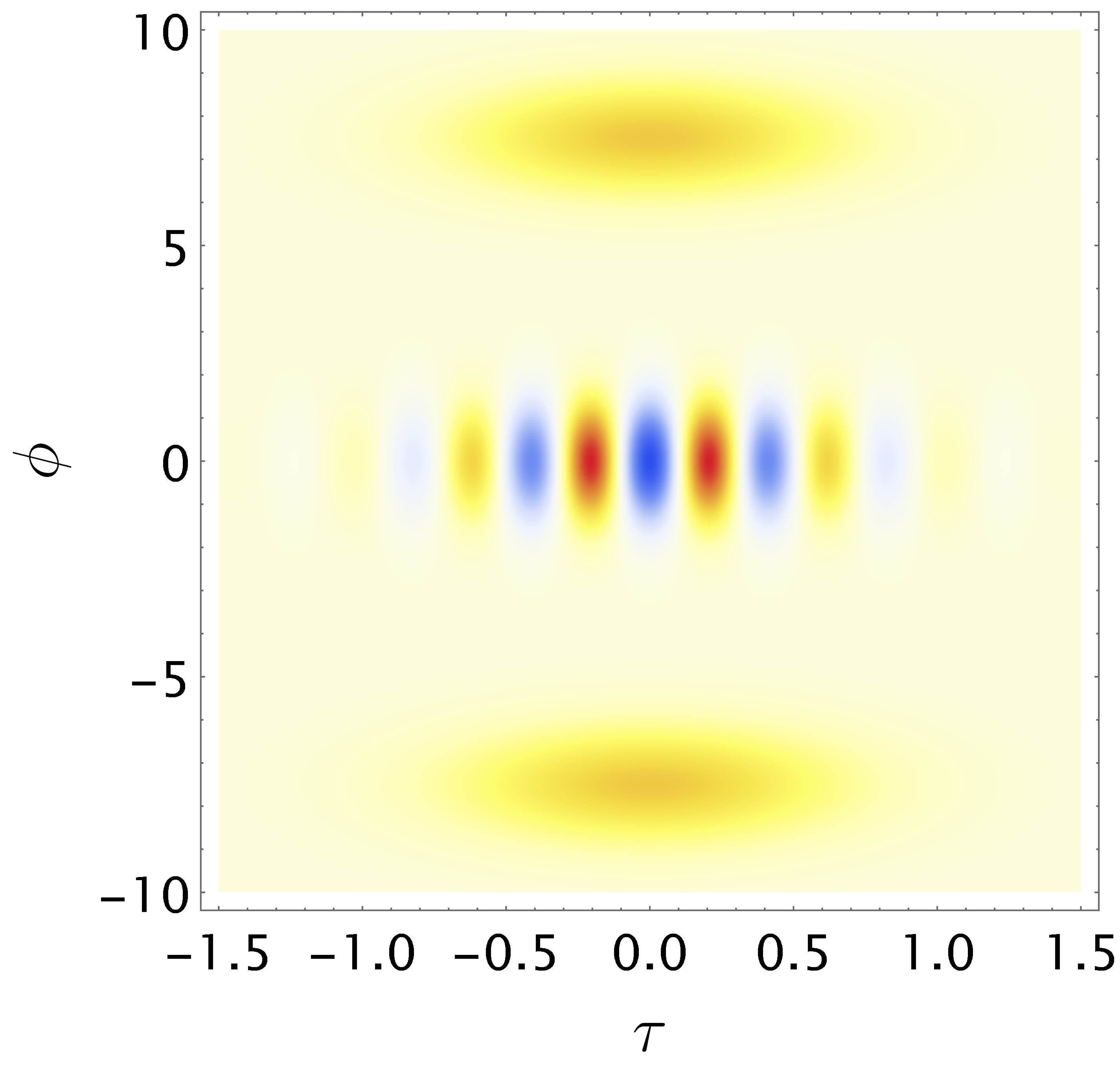}} 
   & \subfloat[Wigner function associated to the collective variable $\Omega$ for a Schr\"odinger cat-like state and $n=2$ photons maximally correlated and saturating the Heisenberg limit. ]{\includegraphics[width=4.5cm]{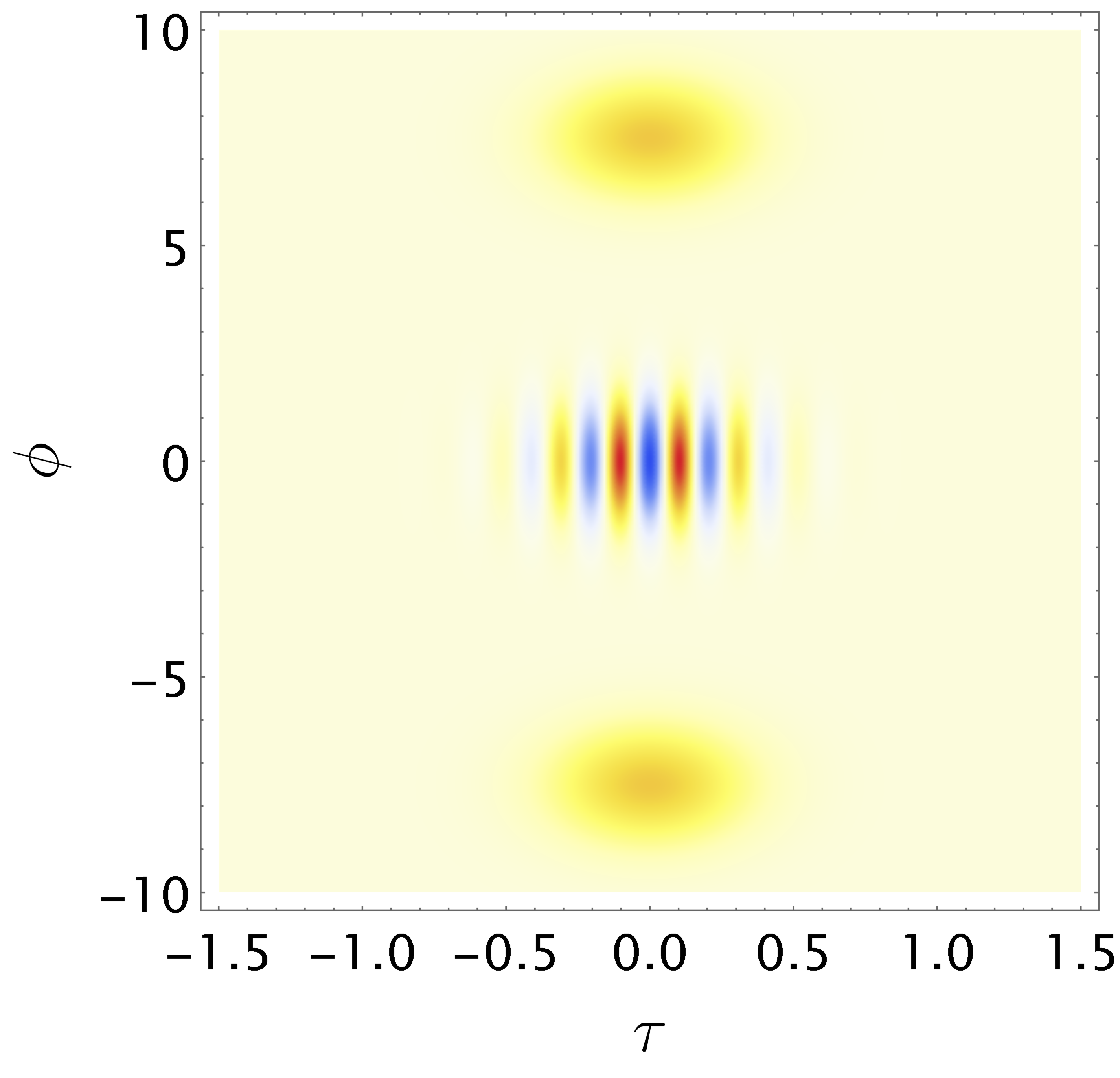}}
&\subfloat[Wigner function associated to the collective variable $\Omega$ for a Schr\"odinger cat-like state and $n=10$ photons maximally correlated and saturating the Heisenberg limit.]{\includegraphics[width=4.5cm]{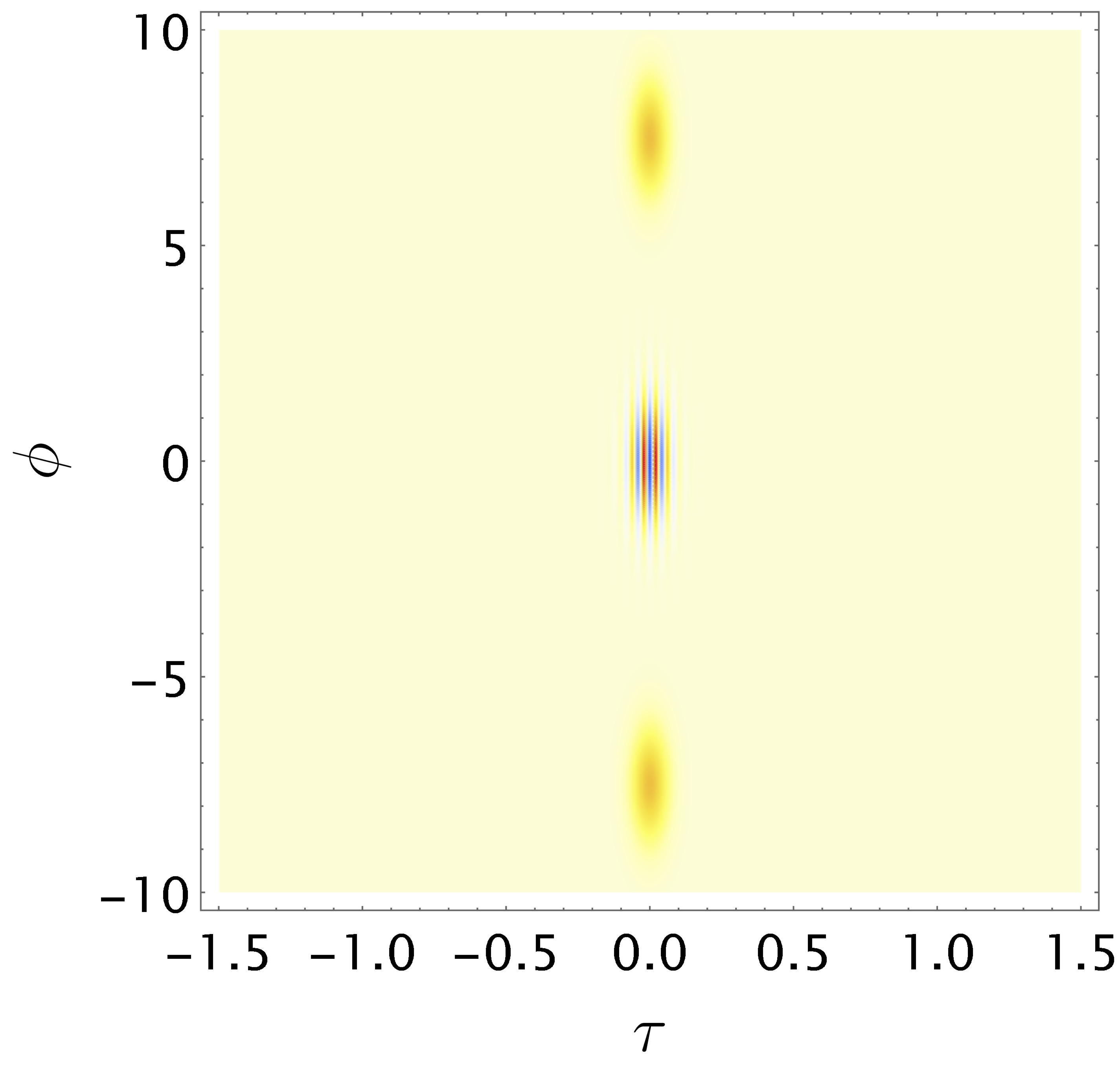}}

\end{tabular}

\caption{}
\label{chats}
\end{figure}

We see in Figure \ref{chats}(a) the Wigner function associated to variable $\omega_1$ (by setting all the other variables to zero) for a Schr\"odinger cat like state as the one shown in \eqref{Ursin}. From this figure, it is clear that the maximal precision provided from this state corresponds to the inter-fringe spacing, which is itself proportional to the distance between the two frequency peaks. However, by depicting this state in the collective variable $\omega_- = \omega_2-\omega_1$ (Fig. \ref{chats}(b), we see that the fringe interspacing scales as $n$, and displacements in time in the phase space associated to this variable can be measured with higher precision. Finally, we show as well the Wigner function associated to a Schr\"odinger cat like maximally correlated state with $n=10$ photons in an (arbitrary)  collective variable in Figure \ref{chats}(c). It's important to recall that even though, in the single photon case, the two-peaked spectrum is a classical interpretation, in the case of $2$ or more entangled photons it is related to the type of mode entangled state, playing thus a role in the quantum properties of the $n$ photon state. 

\section{D. An example comparing our results to a recent experiment and how to improve it}

We now discuss a recent experiment attaining the QCR bound using the Hong-Ou-Mandel (HOM) interferometer and a frequency entangled state. In \cite{chen_hong-ou-mandel_2019}, the authors use a state in the form
\begin{equation}\label{Ursin}
\ket{\psi}=\int d\Omega f(\Omega) (\ket{\omega_1^0+\Omega}\ket{\omega_2^0-\Omega}-\ket{\omega_2^0+\Omega}\ket{\omega_1^0-\Omega}),
\end{equation}
as a probe, and implement a time delay $\delta t$ in one arm of a Hong-Ou-Mandel interferometer. In \eqref{Ursin}, $\omega_i^0$, $i=1,2$, represent the frequency spacing of two well separated center frequency bins.  This time delay is implemented by the evolution operator $e^{i\hat \omega_1 \delta t}$. We notice that state \eqref{Ursin} has precisely the same form as the one found for states displaying a Heisenberg scaling. Moreover, it is as anti-symmetric state, so for $\delta t=0$ it leads to a $100\%$ coincidence probability in the HOM interferometer. Thus, as observed in \cite{fabre_parameter_2021, Jordan2, Lundeen}, it is possible to achieve the QFI using this type of experiment. Moreover, for the considered state, the phase associated to operator $\hat \omega_1$ can be re-expressed as $\omega_-= \Delta+ 2 \Omega$, with $\Delta= \omega_1^0-\omega_2^0$, as in Eq. (3) of the studied reference.  As we can see, this is an evolution that acts only in one photon of the photon pair, so there is no collective effect to be expected. Indeed, we can compute the variance of the generator of the evolution, $\Delta \hat \omega_1 = \Delta^2 +4\sigma_-^2$, where $\sigma_-$ is the root mean square width of the generated photons. Even though the authors do not find exactly this same value for the associated QFI (see also \cite{fabre_parameter_2021}), it is clear that their measurement strategy enables reaching the QCR bound. Nevertheless, this is not the best this experiment can provide, since as presented, it doesn't make use of any collective effect coming from the existing frequency correlation between the photon pair. As a matter of fact, it is possible to implement the dynamical evolution generated by the collective operator $\hat \Omega = \hat \omega_1 - \hat \omega_2$ in this experimental set-up by adding a delay $-\delta t$ in arm $2$ of the interferometer. Consequently, the phase factor would be multiplied by 2, and the associated variance would become $\Delta \hat \omega_- = 4 \Delta^2 +16\sigma_-^2$, which has the predicted $n^2$ scaling ($n=2$ in the present case). 

It is interesting to recall that the HOM experiment is the direct measurement of the Wigner function associated to the variables $\omega_-$ of the biphoton \cite{douce_direct_2013}. Using this result, we can interpret the interference fringes of \cite{chen_hong-ou-mandel_2019} as the interference fringes of a Schr\"odinger cat-like state in the TFPS associated to frequency $\omega_1$. By adding a phase factor in arm 2, these fringes will oscillate two time faster, since in this case, we'll be considering the operator $\hat \omega_-$ as the generator of the dynamical evolution (see also Figure \ref{chats} in Appendix C and the discussion therein). This simple modification to the experiment \cite{chen_hong-ou-mandel_2019} would be a demonstration of the TFPS signature of the Heisenberg scaling and of the results presented in the main text.

\section{E. Discussion about experiments}

We start by discussing the experimental values of the frequency width of the spectral distribution of current photon pair sources, as for instance \cite{chen_hong-ou-mandel_2019, Qudits}. We then mention promising experimental techniques to create larger frequency entangled single photon states.\\

According to the conditions exposed in the main text and in Appendix B, the states displaying a Heisenberg-like scaling for time estimation are frequency correlated or anti-correlated. We discussed in Appendix B a family of frequency-correlated entangled states ($\alpha_i=1 ~ \forall ~i$), while in Appendix D we considered an anti-correlated one with $n=2$, and $\alpha_1=-\alpha_2=1$. Different set-ups, with different physical properties determining the spectral width can produce these two families of states, ant-correlated and correlated ones, and the RMS $\Delta$ will then refer to one type of variable or the other. We must keep this point in mind since we want to evaluate the parameter $\eta$ defined by the relation $\sigma^{2}=(1-\eta)\Delta^{2}$, where $\Delta\gg \sigma$. For simplicity, we'll discuss examples that respect the condition:  $\text{cov}(\omega_{i},\omega_{j})=(\Delta\omega)^{2}=(\text{Var}(\omega_{s}))^{2}$ (a condition that can be met in practice).\\

(1) In \cite{Qudits} the authors use an integrated AlGaAs non-linear optical waveguide working at room temperature producing telecom frequency entangled photon pairs. The width of the joint spectral amplitude along the $\omega_{+}$ axis corresponds to the frequency width of the pump, and it is given by $\sigma=2\pi \times 100$ kHz. As for the width along the $\omega_{-}$ axis, it corresponds to $\Delta=2\pi \times 10.9$ THz. Thus, we find that $\eta \sim 1$, and we have indeed the Heisenberg scaling for all practical values of $n$, since $n\ll \eta/(1-\eta)$. \\
(2) In \cite{chen_hong-ou-mandel_2019}, photon pairs are generated by a bulk ppKTP non-linear crystal, where the frequency width of the phase-matching can be controlled by changing the temperature of the crystal.  The ratio between the spectral width of two peaks in the Joint Spectral Amplitude (JSA) along the $\omega_{-}$ axis and the frequency width along the $\omega_{+}$ axis can be set to 68. This corresponds to $\eta=0.9998$, and the Heisenberg scaling for the estimation of a temporal parameter can be reached for $n \lesssim 4999$.    \\

We now discuss experimental methods to entangle many single photons (more than two, that can be generated by a non-linear interaction). A first solution is to use consecutive non-linear crystals, but its scaling is limited with current technology. 

 The second way is to use existing multiple independent photon sources \cite{Sampling} and entangle them. This can be done by mediating the interaction between two single photons using a third auxiliary one. A potential candidate has been proposed and experimentally demonstrated in \cite{jeannic_dynamical_2021,jeannic_experimental_2021}, where two initially separable single photons in the same pulse interact with a quantum dot embedded into a waveguide, giving rise to an entangled photon pair. The resulting time-of-arrival probability distribution of the photon pairs has an elliptical shape oriented at 45 degrees, indicating the temporal correlation of the photon pairs or, equivalently, their spectral anti-correlation.  Note that the spectral (or temporal) entanglement of the photon pair generated by a spontaneous parametric down-conversion is reproduced with this device. The value of $\eta$ is found to be $\eta=0.99$, as the ratio $\sigma/\Delta\sim 10$, which confirms the relevance of such  experimental device for reaching the Heisenberg scaling in temporal estimation. Once two single photons are entangled, the generation of a larger entangled state could be performed by entangling a third photon with the second, and so on. Multi-photon polarisation entangled states were generated using feedforward and multiplexing \cite{meyer-scott_scalable_2022}. Such an experimental technique is also a potentially good candidate for generating multi-photon frequency entangled states. Developments on the formalisation of the physical mechanism behind frequency entanglement of single photons was recently proposed in \cite{alushi_waveguide_2023}

Finally, note that in practice, we can only measure an unbiased estimator - a statistical estimator that has an expected value which is equal to the true value of the parameter being estimated. If the variance of an unbiased estimator of a parameter attains the Cramér-Rao bound, then the estimator is said to be efficient. This means that the estimator is able to extract as much information about the parameter as possible from the data, and that no other unbiased estimator can have a lower variance. In practice, it is often difficult to find estimators that attain the Cramér-Rao bound, but the bound can be used as a benchmark to compare the performance of different estimators. For instance, in two-photon metrological scenario  \cite{chen_hong-ou-mandel_2019, lyons_attosecond-resolution_2018}, it was shown that the maximum likelihood estimator is efficient for a large number of measurements. 

\section{F. Derivation of Eq. (2) of the main text}

If a state is intrinsically a $n$ mode state it can be mode entangled ({\it i.e.}, diagonal in an entangled mode basis involving different parties) or separable. We'll suppose in this section that we have a $n$ mode separable state. 

The QFI associated to an evolution generated by a Hamiltonian proportional to  $\hat \Omega$ for a state that is separable in $n$ auxiliary modes can be derived as follows: by definition, we have that $\hat \Omega =  \sum_i^n \int {\rm d}\omega \omega \hat a_i^{\dagger}(\omega) \hat a_i (\omega)$, where $i=1,...n$ are the different auxiliary modes. Since the state is mode separable, it is convenient to express this operator in terms of the creation and anihilation operators associated to the $n$ independent modes $\hat b_i$ that diagonalize the coherence matrix ($\langle \hat b_i \hat b_j \rangle= \delta_{i,j}$) with $\hat b_i = \int d\omega S_i(\omega)\hat a_i(\omega)$. Consequently, since we have that 
\beq
\hat \Omega =  \sum_{i,j}^n  \int {\rm d}\omega \omega S_i^*(\omega) S_{j}(\omega) \hat b_i^{\dagger} \hat b_{j}
\eeq
and 
\beq
\hat \Omega^2 =  \sum_{i,j,k,s}^n \int \int{\rm d}\omega {\rm d}\omega' \omega \omega' S_i^*(\omega) S_{j}(\omega)S_k^*(\omega') S_{s}(\omega') \hat b_i^{\dagger} \hat b_{j}\hat b_k^{\dagger} \hat b_{s},
\eeq
so 
\beq
\langle \hat \Omega \rangle= \sum_i^n \sum_j^n \int {\rm d}\omega \omega S_i^*(\omega) S_{j}(\omega)\langle  \hat b_i^{\dagger} \hat b_{j}\rangle = \sum_i \overline \omega_i
\eeq
(since modes $\hat b_i$  diagonalize the coherence matrix)
and
\beq
\langle \hat \Omega^2 \rangle=  \sum_{i,j,k,s}^n \int \int{\rm d}\omega {\rm d}\omega' \omega \omega' S_i^*(\omega) S_{j}(\omega)S_k^*(\omega') S_{s}(\omega') \langle \hat b_i^{\dagger} \hat b_{j}\hat b_k^{\dagger} \hat b_{s}\rangle.
\eeq

In the present Letter, we consider that we're in the narrowband approximation so $ \int {\rm d}\omega \omega S_i^*(\omega) S_{j}(\omega) \simeq \overline \omega \delta_{i,j}$. Thus, $\langle \hat \Omega^2 \rangle$ can be computed as follows: 

\begin{eqnarray}\label{chut}
&&\langle \hat \Omega^2 \rangle=  \sum_{i,j,k,s}^n \int \int{\rm d}\omega {\rm d}\omega' \omega \omega' S_i^*(\omega) S_{j}(\omega)S_k^*(\omega') S_{s}(\omega') \langle \hat b_i^{\dagger} \hat b_{j}\hat b_k^{\dagger} \hat b_{s}\rangle = \nonumber \\
&&\sum_{i,j,k,s}^n \int \int{\rm d}\omega {\rm d}\omega' \omega \omega' S_i^*(\omega) S_{j}(\omega)S_k^*(\omega') S_{s}(\omega') (\langle \hat b_i^{\dagger} \hat \hat b_k^{\dagger}b_{j} \hat b_{s}\rangle + \langle \hat b_i^{\dagger}\hat b_s\rangle \delta_{k,j}).
\end{eqnarray}
For the first r.h.s. term in \eqref{chut} we can use the narrowband approximation, and for the second, we use the fact that the coherence matrix is diagonal and that $\sum_{k,j} \delta_{k,j}S_k^*(\omega') S_{j}(\omega) = \delta(\omega-\omega')$, which leads to 

\beq
\langle \hat \Omega^2 \rangle= \sum_{i,s}^n \overline \omega_i \overline \omega_s \langle \hat b_i^{\dagger} \hat \hat b_s^{\dagger}b_{i} \hat b_{s}\rangle + \sum_{i}^n \overline{\omega^2}\langle \hat b_i^{\dagger}\hat b_i\rangle.
\eeq

We can now compute $(\Delta \hat \Omega)^2 = \langle \hat \Omega^2 \rangle - \langle \hat \Omega \rangle^2$ by adding and subtracting the term $\sum_i^n \overline \omega_i^2 \langle \hat b_i^{\dagger}\hat b_i\rangle$ to the whole expression, and recalling that  the QFI  for pure states is proportional to this variance for the type of evolution considered. Hence, under the separability assumption, we obtain Eq. (2) of the main text:

\begin{equation}\label{variancesqueeze}
(\Delta \hat \Omega)^2 =  \sum_i^n \langle \hat n_i \rangle  \Delta \omega_i+{\overline \omega_i}^2 (\Delta \hat n_i)^2.
\end{equation}

\bibliography{biblioMetro}

\begin{thebibliography}{60}
\bibitem [{Com()}]{Comment}%
  \BibitemOpen
  \href@noop {} {\bibinfo {title} {Notice that we can associate $\overline
  {\omega^2}$ to the variance of a spectral distribution centered at $\bar
  \omega = 0$.}}\BibitemShut {Stop}%
\end{thebibliography}

\end{document}